\documentclass[1pt,a4paper,nobibnotes,nofootinbib]{revtex4}

\usepackage[english]{babel}
\usepackage{tabularx}

\usepackage{amssymb}
\usepackage{amsmath}
\usepackage{epsfig}
\newcommand{\be}{\begin{equation}}
\newcommand{\ee}{\end{equation}}
\newcommand{\bea}{\begin{eqnarray}}
\newcommand{\eea}{\end{eqnarray}}

\newcommand\lr[1]{{\left({#1}\right)}}
\newcommand{\egamma}{E_{\gamma}}
\newcommand{\units}[1]{\,\hbox{#1}}
\newcommand{\unitMass}{\,\units{GeV/c}^2}

\newcommand{\GeV}{\units{GeV}}
\newcommand{\TeV}{\units{TeV}}
\newcommand{\fb}{\units{fb}}

\newcommand{\invpb}{\units{pb}^{-1}}
\newcommand{\invfb}{\units{fb}^{-1}}
\newcommand{\lumi}{$\mathcal{L}$}
\newcommand{\lumiunit}{\,\hbox{cm}^{-2}\hbox{s}^{-1}}
\newcommand{\pomeron}{\mathbb{P}}

\newcommand{\dkap}{\Delta\kappa^{\gamma}}

\newcommand{\lam}{\lambda^{\gamma}}
\newcommand{\wwgamma}{WW\gamma}
\renewcommand{\d}{\mathrm{d}}
\newlength{\picwidth}
\begin{document}
\title{Anomalous WW$\gamma$ coupling in photon-induced processes using forward detectors at the LHC}
\author{O. Kepka}\email{kepkao@fzu.cz}
\affiliation{CEA/IRFU/Service de physique des particules, CEA/Saclay, 91191
Gif-sur-Yvette cedex, France}
\affiliation{IPNP, Faculty of Mathematics and Physics,
Charles University, Prague} 
\affiliation{Center for Particle Physics, Institute of Physics, Academy of Science, Prague} 
\author{C. Royon}\email{royon@hep.saclay.cea.fr}
\affiliation{CEA/IRFU/Service de physique des particules, CEA/Saclay, 91191 
Gif-sur-Yvette cedex, France}

\begin{abstract}
We present a new method to test the Standard Model expectations at the LHC
using photon-induced $WW$ production. Both $W$ decay in the main ATLAS or CMS
detectors while scattered protons are measured in forward detectors. 
The sensitivity to anomalous $WW\gamma$ triple gauge coupling can be improved by
more than a factor 5 or 30 compared to the present LEP or Tevatron sensitivity
respectively. 
\end{abstract}
\maketitle

\section{Introduction}
\par In this paper, we discuss a new possible test of the Standard Model (SM) predictions
using photon induced processes at the LHC, and especially
the $WW$ production. The cross sections of these
processes are computed with high precision using Quantum Electrodynamics
(QED) calculations, and an experimental observation leading to differences with
expectations would be a signal due to beyond standard model effects. The
experimental signature of such processes is the decay products of the $W$ in the
main central detectors from the ATLAS and CMS experiments and the presence of
two intact scattered protons in the final state.

\par It is foreseen to equip two LHC experiments, ATLAS and CMS, with very
forward detectors which allow to detect intact scattered protons 
at very small angles after the collision.
Together with the main central detector, they
will help to identify inclusive and exclusive diffractive processes, two-photon
exchange, etc. that would otherwise elude to be seen in the standard way with
tproduction he central detector only. In this paper, we extend the physics diffractive program
at the LHC by discussing $WW$ production via photon induced processes.
Two-photon interactions show a very
clear experimental signature: two protons detected in the forward detectors and
a certain number of well isolated leptons in the central (ATLAS or CMS) detector
when at least one of the $W$ decays leptonically.
The activity in the central detector is devoid of any other
particles since the interaction is due to the colorless exchange of two photons.

\par The theoretical success of the electroweak part of the SM lies in
prescribing the underlying SU(2)$\times$U(1) symmetry to the fermion fields
which not only implies the form of the coupling between fermions and gauge fields,
but also leads to non-trivial predictions on interaction between the gauge
fields themselves. It is quite difficult to measure the boson self-couplings
because only the decay products of the bosons are observed and 
several different bosonic vertices can contribute to one observable process.
The International Linear Collider is believed to be the machine for high
precision measurement of the boson couplings but significant improvement of our
knowledge of the boson self-interaction can be achieved already at the LHC. 

\par The $W$ pair production via two-photon exchanges shows a sufficiently high cross section
to be observed at the LHC
and a few thousand of such events can be observed in three years of LHC
running at low luminosity. Using these data, one can directly measure the two-photon $W$ pair
production cross section and constrain the triple gauge coupling (TGC)
$\wwgamma$ and quartic gauge coupling (QGC) 
$WW\gamma\gamma$ in  $pp\rightarrow pWWp$ processes through $\gamma\gamma\rightarrow WW$.
Potential deviations from the SM expectation could indicate new physics beyond the SM.
In this report we will show how the triple gauge boson vertex
$\wwgamma$ can be constrained using these topologies, by tagging
the protons in forward detectors and observing the decay products of the $W$ in
the central detector. 

\par The outline of this paper is as follows. We start by describing
the forward detectors used in this measurement in Section 2.
The concept of two-photon production will be overviewed in Sections
3 and 4 with the emphasis on the $W$ pair production. In Section 5, we describe a
measurement of the two-photon $WW$ production cross section using forward
detectors and discuss the background and resolution issues. Section 6 introduces
the anomalous parameters for the $WW\gamma$ vertex into the Lagrangian and
derive the sensitivity on those couplings that could be achieved at the LHC using
forward detectors. 

\section{Forward detectors at the LHC}
Following the experience from HERA~\cite{h1zeus} and the
Tevatron~\cite{cdf} new detectors that can operate at the highest LHC
luminosities
are proposed to be installed in the LHC tunnel as an additional upgrade of the
ATLAS and CMS detectors.  These magnetic spectrometers measure precisely very
forward protons and allow to study SM two-photon and diffractive physics such
as $W$ or jet production, search for the SM Higgs boson and also for new
physics signals such as SUSY, in conjunction with the corresponding central
detectors~\cite{higgs,survivalKhoze,gammaSUSY}. Protons which do not break up
during collisions loose a small fraction of their momentum, are scattered at
very small angles ($<100\units{mrad}$) and continue to travel down the beam
pipe. The bending magnets deflect them out of the beam
envelope allowing the proton detection. Specifically for ATLAS and CMS, it is
proposed to
detect the protons
at a distance of 220\units{m} and 420\units{m} from the ATLAS/CMS interaction point.
Proton tracks can be reconstructed from hits in several layers of
silicon 3D detectors that will approach the beam as close as 2\units{mm} and
5\units{mm} for 220\units{m} and 420\units{m} stations, respectively~\cite{RP220dis,FP420rnd}.

\par A prime process of interest to be studied with forward detectors is the
Central Exclusive Production (CEP)  $pp\rightarrow p +\phi+p$, in which a
single particle $\phi$ such as Higgs boson or other (predominantly scalar)
particle  may be created in certain physics
scenarios~\cite{higgs,survivalKhoze,ww}.  The observation of the new particle
would allow a direct determination of its
quantum numbers and a very precise determination of the mass, irrespective of
the decay channel of the particle with a resolution between 2$\unitMass$ and
3$\unitMass$ per event~\cite{RP220dis,FP420rnd}. Exclusive production is a new phenomenon which was
observed recently for the first time in the measurements of the CDF 
collaboration~\cite{CDFexcDijets,exclusiveTwoGammaCDF,CDFexcChic,us,chic} and the confirmation and exploration of 
the CEP is a prime goal of the forward physics program, especially concerning the Higgs 
boson \cite{CEPhiggs} and SUSY particle production~\cite{diffSusy1}.
\section{Photon-induced processes at the LHC}

\par Another important physics application of the forward detectors at the LHC 
are the photon-induced interactions. Processes in
which at least one of the colliding particles emits almost real photons that
subsequently enter the hard interaction have been already well
explored at the electron-positron and proton-electron colliders at LEP and
HERA, respectively. Very recently, photon-photon and photon-proton processes
were also measured at a hadron-hadron collider by the CDF
collaboration. In particular, CDF recorded isolated electron-positron pairs
\cite{CDFdileptons} with large rapidity gaps produced in $pp\rightarrow
pl^+l^-p$ through $\gamma\gamma\rightarrow l^+l^-$ and also $\Upsilon$ candidates in
diffractive photoproduction $pp\rightarrow p\Upsilon p$ through $\gamma\pomeron
\rightarrow\Upsilon$~\cite{CDFupsilon}. The obtained agreement between
the two-photon dilepton
production cross section measurement with the theoretical prediction proved that 
the definition of exclusive process at CDF was well understood and could 
in turn be applied for the exclusive two photon production~\cite{exclusiveTwoGammaCDF}.
\par As it was reviewed in~\cite{FP420rnd}, the LHC program of
photon-induced interactions includes the two-photon
production of lepton pairs that will be used for the independent luminosity
measurement,
two-photon production of $W$ and $Z$ pairs as a mean to
investigate anomalous triple and quartic gauge couplings, two-photon production
of supersymmetric pairs, associated $WH$ photoproduction, and anomalous single
top photoproduction. Last but not least, the dimuon two-photon production will
be used for calibration and an independent alignment of the forward detectors at
420\units{m} with respect to the beam on a store-by-store basis. During 
the low luminosity runs at the early stage of the LHC, it will be possible 
to identify two-photon processes without forward detectors. In this case
one relies on the observation of the large rapidity gaps between the
centrally created object and the forward protons. On the other hand, 
in order to run also at high instantaneous luminosity \lumi$=10^{33}-10^{34}\lumiunit$
when up to 35 interaction per bunch crossing will occur, the operation of forward
detectors is inevitable to distinguish whether the tagged proton come from
the main $\gamma \gamma$ interaction and not from a pileup event.
In addition, the operation of forward detectors allows to reconstruct precisely the mass
of the central object using the missing mass method as $M=\sqrt{s\xi_1\xi_2}$
\cite{missingmass}, where $\xi_i$ is the momentum fraction loss of the protons
$\xi=(|\vec{p}|-|\vec{p}'|)/|\vec{p}|$.  The acceptance of the ATLAS forward
detectors proposed by the ATLAS Forward Physics (AFP) Collaboration spans
$0.0015<\xi<0.15$ allowing the
detection of object with masses $100\units{\GeV}<M<800\units{\GeV}$ with a good efficiency.
The acceptance in CMS is similar.

\begin{figure}[t]
\includegraphics[width=.8\picwidth]{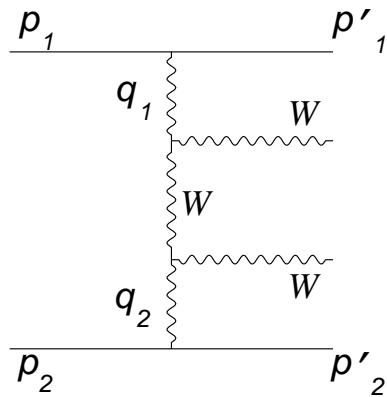}
\caption{S-channel diagram, one of the three Feynman diagrams of the $W$ pair production in two-photon 
 processes. The others not shown correspond to a u-channel diagram and a diagram with 
direct coupling $\gamma\gamma WW$. $p'_i$ are protons that do not break up but continue
to travel down the beam pipe at small angles.}
\label{fig:diagram}
\end{figure}

\section{Two-photon $W$ pair production}

\par
\begin{figure}[htb]
\includegraphics[width=1\picwidth]{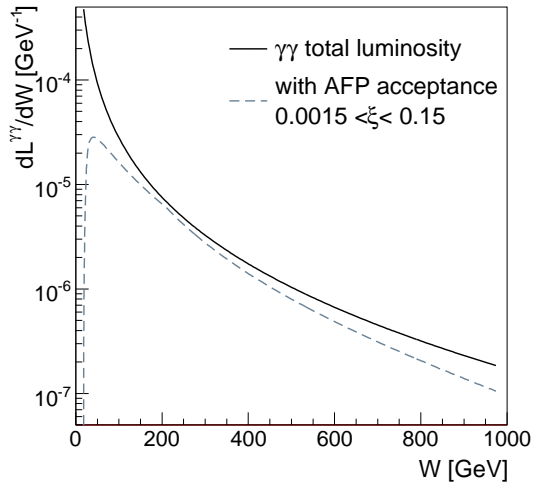}
\caption{Relative effective $\gamma\gamma$ luminosity in $pp$ collisions at $14\units{TeV}$
as a function of the  two-photon invariant mass. The maximal virtualities of the
emitted photons are set to $Q^2_{max}=4\,\GeV^2$. The dashed curve
shows the photon spectrum within the ATLAS or CMS forward detector acceptance.}
\label{fig:luminosity}
\end{figure}

 The two-photon production is described in the framework of Equivalent Photon Approximation 
(EPA)~\cite{Budnev}. The almost real photons (low photon virtuality $Q^2=-q^2$) are 
emitted by the incoming protons producing an object $X$, $pp\rightarrow pXp$, 
through two-photon exchange $\gamma\gamma\rightarrow X$, see Fig.~\ref{fig:diagram}.
The photon spectrum of virtuality $Q^2$ and
energy $E_{\gamma}$ then reads~\cite{Budnev}
\begin{equation}
\d N = \frac{\alpha}{\pi}\frac{\d E_{\gamma}}{E_{\gamma}}\frac{\d Q^2}{Q^2}
 	 \left[ \left(1-\frac{E_{\gamma}}{E}\right)\left(1-\frac{Q^2_{min}}{Q^2}\right)F_E +
	         \frac{E_{\gamma}^2}{2E^2}F_M\right],
\label{eq:flux}
\end{equation}
where $E$ is the energy of the incoming proton of a mass $m_p$, $Q^2_{min}\equiv
m^2_p E_{\gamma}/[E(E-E_{\gamma})]$ the photon minimum virtuality allowed by
kinematics and $F_E$ and $F_M$ are functions of the electric and magnetic form factors. They read
in the dipole approximation~\cite{Budnev}
\begin{equation}
F_M=G^2_M,  \quad F_E=(4m_p^2G^2_E+Q^2G^2_M)/(4m_p^2+Q^2),\quad G^2_E=G^2_M/\mu_p^2=(1+Q^2/Q^2_0)^{-4}.
\end{equation}
The magnetic moment of the proton is $\mu_p^2=7.78$ and the scale $Q^2_0=0.71\,\GeV^2$.
The photon flux function falls rapidly as a function of the photon energy $E_{\gamma}$ 
which implies that the two-photon production dominates the low mass region
of the produced system $W\approx2\sqrt{E_{\gamma1}E_{\gamma2}}$. After integrating the product of the
photon fluxes from both protons over the photon virtualities and the energies while keeping
the two-photon invariant mass fixed to $W$, one obtains the two-photon effective
relative luminosity spectrum\footnote{We thank K. Piotrzkowski for notifying us
about the sign error in the photon spectrum formula published in~\cite{Budnev}. For that reason
we present the correct flux 
formula in Appendix (see Formula \ref{app:eq:budnev}) for the reader's convenience.}.  The 
effective $\gamma \gamma$ luminosity at low $W$ energy
is shown in Fig.~\ref{fig:luminosity} in full line. The luminosity spectrum was
calculated using the upper virtuality bound $Q^2_{max}=4\,\GeV^2$. The luminosity spectrum 
after taking
into account the forward detector acceptances
$0.0015<\xi<0.15$ is also shown in the figure in dashed line
(it is calculated in the limit of low
$Q^2$, thus setting $E_{\gamma}=\xi E_{beam}$). Using the effective relative photon
luminosity $\d L^{\gamma\gamma} \slash \d W$, the total cross section for a certain sub-process
reads 
\begin{equation}
      \frac{\d\sigma}{\d\Omega}=\int\frac{\d \sigma_{\gamma\gamma\rightarrow X}(W)}{\d\Omega}\frac{\d L^{\gamma\gamma}}{\d W}\d W, 
\label{eq:totcross}
\end{equation}
in which the $\d \sigma_{\gamma\gamma\rightarrow X}/\d\Omega$ denotes the differential 
cross section of the sub-process $\gamma\gamma\rightarrow X$ that is a function of the 
invariant mass of the two-photon system. Finally, after multiplying the luminosity spectrum by
the luminosity of the machine one obtains the two-photon production cross section.

\par The $\gamma\gamma \rightarrow WW$ pair production is composed of three
distinct Feynman diagrams (one of them is shown in Fig.~\ref{fig:diagram}).
The complete formula of the differential cross section is listed in
Appendix, see formula (\ref{app:eq:wwprod}). Using formulae (\ref{eq:totcross}) 
and (\ref{app:eq:wwprod})
and integrating over the two-photon mass $W$ and spatial angles $\Omega$, one
obtains the total cross section for $W$ pair production through $\gamma\gamma$
exchange in $pp$ collision which is $\sigma_{WW}=95.6\units{fb}$. This number was 
obtained for photons with an upper virtuality of $Q^2_{max}=4$\,GeV$^2$ and
photon energies $0<E_{\gamma1,2}<$7\units{TeV}.  No QED survival probability
(to be discussed below) was taken into account.

\par Several extensions of the SM predict new physics at the TeV scale 
(e.g. low energy supersymmetry at about 1\TeV, new strong dynamics at about
1\TeV, low energy scale (quantum) gravity and extra large dimensions at a few
TeVs). All models give a solution to the so called hierarchy problem, namely the question
why the Plank mass $M_{\mathrm{Plank}}\simeq10^{19}\GeV$ is huge compared to
the electroweak scale $M_\mathrm{EW}\simeq M_Z$. 
The observation and the measurements of photon induced processes with a photon invariant mass $W>1\TeV$ 
is thus particularly interesting.
The total production cross section of the $W$ pairs with the invariant mass $1\TeV<W<14\TeV$ via two photon exchange
is $\sigma_{WW}(W>1\TeV)=5.9\fb$. After taking into account the acceptance of the ATLAS or CMS forward 
detectors, the 
cross section falls to  $\sigma_{WW}(W>1\TeV)=2.0\fb$. This means that it is expected
to accumulate about 400 $WW$ events with an invariant mass above 1\TeV with a luminosity of 
\lumi=200$\invfb$.

\section{Measuring the $\gamma\gamma\rightarrow WW$\ Standard Model cross section}

$\gamma\gamma\rightarrow WW$ is a very interesting process 
to measure precisely at the LHC since it incorporates the fundamental SM property of 
diagram cancellation.
The SM model is a renormalizable theory. A necessary condition for the renormalizibility
of the theory into all orders is the so called "tree unitarity" 
that demands that unitarity is only minimally (logarithmically) violated in any fixed order 
of the perturbation series~\cite{TreelevUnit1,TreelevUnit2}.
More precisely the tree level unitarity means that any $n$-point tree level amplitude $M^{n}_{tree}$
of the process $1+2\rightarrow 3+4+\dots+n$ grows for the fixed nonzero angles in the high energy
limit $E\rightarrow\infty$ not faster than $M^{n}_{tree}=O(E^{4-n})$. 
In particular for the binary process of  $W$ pair production, the tree level unitarity implies that the 
scattering amplitude $\gamma\gamma\rightarrow WW$ should turn constant in the high energy
limit.
\par There are two diagrams with $W$ exchange and one diagram with a direct $\gamma\gamma WW$
coupling to be considered. The longitudinal polarization vector of the vector boson of nonzero mass 
grows linearly as a function of the boson momentum. Therefore each of the scattering diagrams 
alone is quadratically divergent in terms of the boson momentum in the high energy limit. However, 
the quadratic divergence is cancelled when all three diagrams are added and
the total cross section is constant in the two-photon $W^2_{\gamma\gamma}$ high
invariant mass limit (the linear divergence arising in the case when only one
of the vector bosons in the final state is longitudinally polarized is cured by
introducing another scalar in the theory, the Higgs boson). It is thus very
important to measure precisely the $\gamma\gamma\rightarrow WW$ 
process since it incorporates the fundamental feature of the SM diagram cancellation.

\par The experimental signature of diboson events is very clear. Depending on the decay
of the $W$ there is zero, one, or two leptons in the final state. When both the $W$ decay
purely hadronically 
four jets are produced in the final state. This topology can be easily  mimicked 
in the high luminosity environment
with pileup interactions and also suffer from a high QCD
background. Therefore this case is not considered in the following, and we always require
that at least one of the $W$ decays leptonically.
In addition, the interpretation of the signal is simple contrary to e.g. $e^{+}e^{-}\rightarrow WW$ production
at LEP where such production could be due to $\gamma$ or $Z$ exchange and one could 
not clearly separate the $\wwgamma$ and $WWZ$ couplings. In our case, only the $\gamma$
exchange is possible since there is no $Z\gamma\gamma$ vertex in the SM.
 
\par In summary, we require the following constraints at particle level to select $WW$ events: 
\begin{itemize}
   \item both protons are tagged in the forward detectors in the acceptance $0.0015<\xi<0.15$
   \item at least one electron or muon is detected with $p_T>30\units{GeV/c}$ and 
    $|\eta|<2.5$ in the main detector
\end{itemize}

\par The main source of background is the $W$ pair production
in Double Pomeron Exchange (DPE), i.e. $pp\rightarrow p+WW+Y+p$ through 
$\pomeron\pomeron\rightarrow WW+Y$
where $Y$ denotes the pomeron remnant system. The rapidity gaps in DPE are
smaller in size than in two-photon production because of the pomeron remnants 
and therefore it should be in principal possible to remove some part of the background 
by the rapidity gap size requirement. But running at high luminosity does not allow
to rely on rapidity gap selection since gaps can be easily spoiled
by particles coming from pile up interactions. 
The two-photon diboson production and the pomeron background were
simulated using the Forward Physics Monte Carlo (FPMC) that was developed to enable 
event generation of all forward physics studies in one framework~\cite{FPMC}.
Besides the mentioned processes, it can simulate
Central Exclusive Production in hadron-hadron collisions, single diffraction, diffraction
at HERA, diffractive dissociation, etc.

\par The diffractive and two-photon production cross sections were multiplied
by the survival probability factor, the probability that both the protons escape
intact and are not destroyed by additional soft exchanges that might occur in addition to the
hard interaction. Since the photon or pomeron induded  processes are of different nature, 
the corresponding survival
probability differs.  It is is by more than one order of magnitude smaller (0.03) for
the diffractive production than for the QED two-photon survival probability
(0.8)~\cite{survivalKhoze}. Note that the survival probability for pomeron
exchanges is model dependent (Ref.~\cite{survivalFrankfurt} and
\cite{survivalMaor}  predict respectively 0.01 and 0.03) and it will be one of the first diffractive measurements
to be performed at the LHC . Hence the uncertainty due
with the DPE background is about a factor of three, and we take
the highest value predicted for the survival probability (0.03) to be conservative,
which leads to the higher background DPE contamination. In summary, taking the 
survival probability factor into account, the total DPE $WW$ production cross section is
of the order of 67\fb.   

\par
\begin{figure}[t]
\begin{center}
\includegraphics[height=1\picwidth]{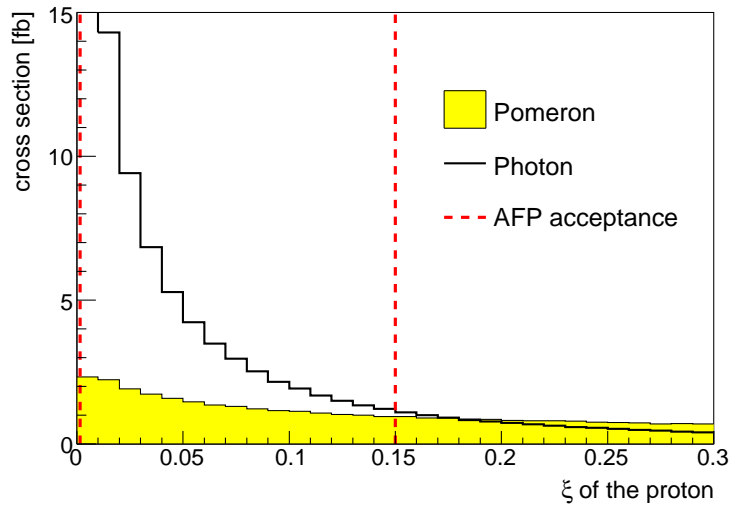}
\end{center}
\caption{
$\xi$ dependence of the cross section for photon (black line) and
pomeron induced events (shaded region). The former shows a steep $\xi$ dependence
while the pomeron
background is suppressed by tagging the protons within the acceptance
$0.0015<\xi<0.15$. 
} 
\label{sbxi}
\end{figure}

 To remove most of the DPE background, it is possible to cut on the $\xi$ of the protons
measured in the proton taggers.
Indeed, two-photon
events populate the low $\xi$ phase space whereas DPE events show a flat $\xi$ distribution,
see Fig.~\ref{sbxi}. The lower proton momentum
fraction $\xi$ we can measure in the data sample within the forward detector acceptance, 
the higher signal (S) to background (B) ratio we obtain. 
A cut on the maximal value of $\xi$ is applied in data to enhance the signal over background ratio.
In table~\ref{tab:sb_vs_xi}, we give the signal and the background cross sections after 
analysis cuts
requiring the presence of a reconstructed lepton (electron or muon) and the tag of both protons in forward detectors
after different cuts on $\xi$.
We also give
the $S/\sqrt{B}$ ratios for an integrated luminosities of $200\invpb$ and $1\invfb$, 
respectively. The $pp\rightarrow pWWp$ cross section can be measured precisely with a \lumi=$1\,\invfb$ 
with a statistical
significance higher than 20$\sigma$ depending on the active $\xi$ range. Using
the full $\xi$ acceptance $0.0015<\xi<0.15$, one expects about 30
tagged $WW$ events. As the upper cut $\xi_{max}$  decreases, one obtains a cleaner
signal, but the number of observed events drops, see Tab. \ref{tab:sb_vs_xi}.
We note that already with a low integrated luminosity of \lumi=$200\,\invpb$ it is 
possible to observe 5.6 $W$ pair two-photon events for a background of DPE lower than 0.4,
leading to a signal 
above 8 $\sigma$ for $WW$ production via photon induced processes. 

\begin{table}[h]
\begin{tabular*}{.6\textwidth}{@{\extracolsep{\fill}}c|cc|lcc}
 $\xi_{max}$&         signal [fb]&     background [fb]   &$S/\sqrt{B}$ &\lumi=$200\invpb$ & 
                                                \lumi=$1\invfb$  \\
\hline
           0.05&        13.8&       0.16&          &     15&              34 \\
           0.10&        24.0&          1.0&        &    11&              24 \\  
           0.15&        28.3&          2.2&        &    8.6&              19 \\
\end{tabular*}
\caption{Signal and background cross sections for $\gamma\gamma\rightarrow WW$ production,
and $S/\sqrt{B}$ ratios for two luminosities ($200\invpb$ and
1$\invfb$) as a function of the forward detector acceptance $0.0015<\xi<\xi_{max}$. 
The presence of at least one reconstructed lepton is required 
as mentionned in the text.}
\label{tab:sb_vs_xi}
\end{table}

\section{Anomalous $\wwgamma$ triple gauge coupling}
\par
\begin{figure}[h]
\includegraphics[width=1\picwidth]{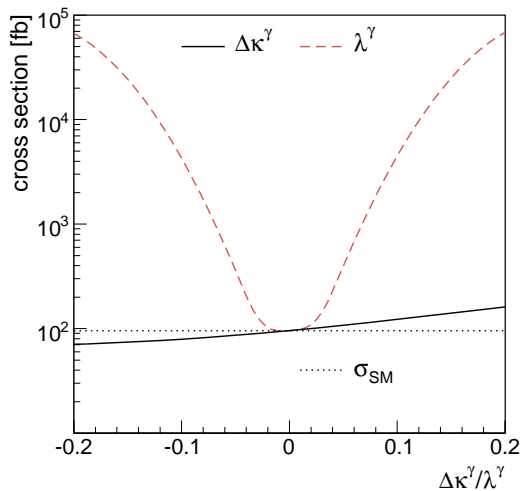}
\caption{The full two-photon $pp\rightarrow pWWp$ production cross section as a
function of the $\wwgamma$ anomalous coupling parameters $\dkap$ and $\lam$ for
$\Lambda=\infty$, i.e. with no form factor assumed. The SM value corresponding to 
$\dkap=\lam=0$ is $\sigma_{SM}=95.6\fb$.}
\label{fig:anomcross}
\end{figure}

 New physics with a characteristic scale (i.e. the typical mass of new particles) 
well above  what can be probed
experimentally at the LHC can manifest itself as a  modification of gauge
boson couplings due to the exchange of new heavy particles. The conventional
way to investigate the sensitivity to the potential new physics is
to introduce an effective
Lagrangian with additional higher
dimensional terms parametrized with anomalous parameters. In this paper, 
we consider the modification of the $\wwgamma$ triple gauge boson vertex with additional terms conserving $C-$ and $P-$
parity separately, that are parametrized with two anomalous parameters $\dkap$, $\lam$ (a similar study discussing
how to constrain the anomalous quartic coupling at the LHC using forward detectors can be found in \cite{QuarticCoupling}).
The effective Lagrangian reads
\begin{equation}
   \mathcal{L}/ig_{WW\gamma}=(W^{\dagger}_{\mu\nu}W^{\mu}A^{\nu}-W_{\mu\nu}W^{\dagger\mu}A^{\nu})
   +(1+{\Delta\kappa^{\gamma}})W_{\mu}^{\dagger}W_{\nu}A^{\mu\nu}+\frac{{\lambda^{\gamma}}}{M_W^2}W^{\dagger}_{\rho\mu}
   W^{\mu}_{\phantom{\mu}\nu}A^{\nu\rho},
\end{equation}
where $g_{WW\gamma}=-e$ is the $\wwgamma$ coupling in the SM and the double-indexed terms are
$ V_{\mu\nu}\equiv\partial_{\mu} V_{\nu}-\partial_{\nu}V_{\mu}\nonumber$, for $V^{\mu}=W^{\mu},A^{\mu}$. In the 
SM, the anomalous parameters are $\dkap=\lam=0$.  The dependence of the total
cross section on the anomalous parameters is shown in Fig.~\ref{fig:anomcross}.
The enhancement of the $WW$ cross section is quite different for both couplings.
It is strong for $\lam$ where the cross section can be enhanced by more than
two orders of magnitude.
It is parabolic, monotonic for $\dkap$ in the interesting region around the SM
values and the dependence on $\dkap$ is very weak. 
The event generation was carried out within the FPMC program~\cite{FPMC} which was
interfaced with the O'Mega matrix element generator~\cite{OMEGA} for this
purpose. O'Mega constructs all possible LO
contributions to particular process and generates a callable fortran function for
the matrix element.

\par As was mentioned before, the SM couplings
are defined in such a way that all tree level amplitudes fulfill the
condition of tree level unitarity. This property is broken when additional terms of higher dimension are
introduced in the Lagrangian.
It is necessary to introduce a cut-off that removes the effect of the anomalous
couplings in that limit in order to restore the tree level unitarity.
Conventionally, this is done by introducing form factors in the dipole form
\begin{eqnarray}
\dkap &\rightarrow& \frac{\dkap}{(1+s_{\gamma\gamma}/\Lambda^2)^n},\nonumber\\
\lam &\rightarrow& \frac{\lam}{(1+s_{\gamma\gamma}/\Lambda^2)^n},
\label{eq:formfactors}
\end{eqnarray} 
where the $\Lambda$ is a cut-off scale, $n$ is an integer number and $s_{\gamma\gamma}$ 
is the squared invariant mass of the photon-photon system. All the following
calculations in which the form factors are taken into account were done setting $n=2$. 
Most of analyses dealing with anomalous couplings 
use either no form factor  $\Lambda=\infty\units{TeV}$ or $\Lambda=2\units{TeV}$. In order
to compare our results with other approaches we will therefore systematically present our limits for
both choices.

\par
\begin{figure}[htb]
\begin{center}
\includegraphics[width=\picwidth]{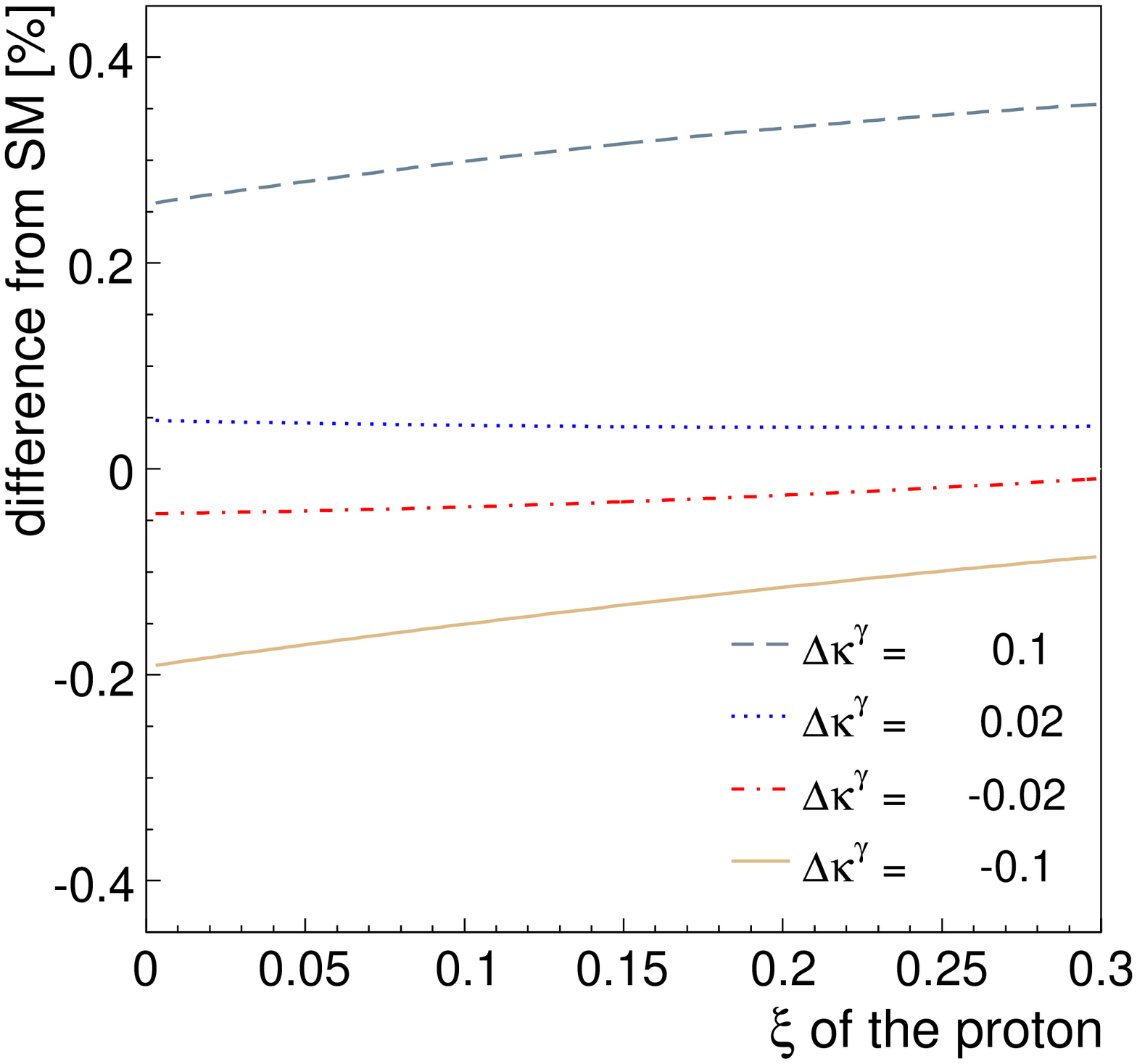}
\includegraphics[width=\picwidth]{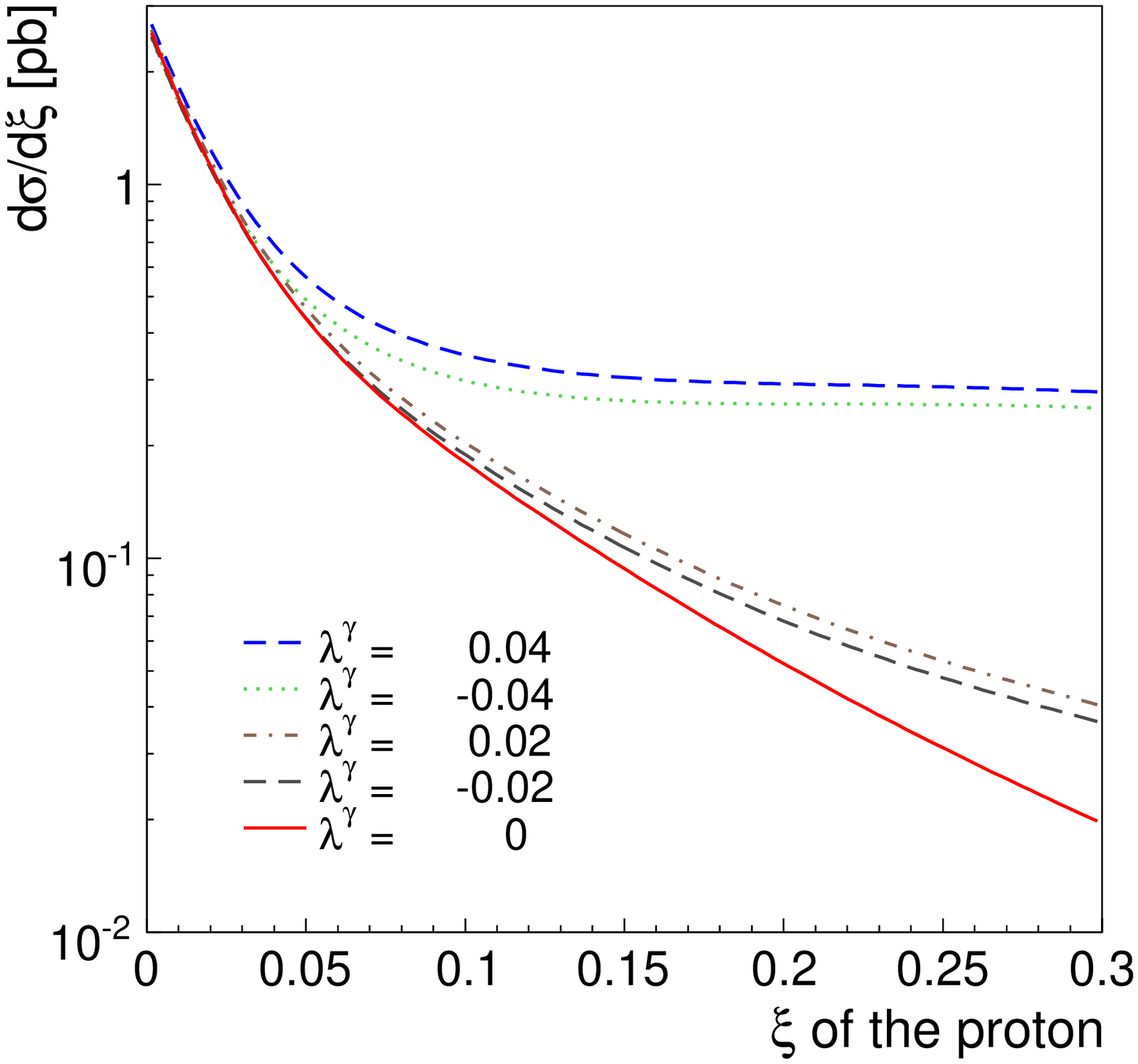}
\end{center}
\caption{$\xi$ dependence of the two-photon $WW$ cross section for different values of 
$\dkap$ (left) and $\lam$ (right) (SM values are 0).
For $\lam$, the cross section is enhanced at high 
$\xi$ which is at the edge of the forward detector acceptance ($\xi=0.15$). 
On the contrary, varying $\dkap$ in the interesting 
range $(-0.05<\dkap<0.05)$ changes mainly the normalization and not the shape of the $\xi$ distribution.}
\label{fig:xianom}
\end{figure}

 The anomalous parameters $\dkap$ and $\lam$ have different effects on various
observables. It turns out that the $\dkap$ changes mainly the normalization
of the distributions whereas the $\lam$ parameters modify the shape of the 
observables, which will be shown in the next section for some 
angular distributions. Here, the differential 
cross section is plotted as a function of the momentum fraction $\xi$ in Fig.~\ref{fig:xianom}.
$\dkap$ enhances the cross section any $\xi$ in contrast to the $\lam$ parameter which 
modifies the high $\xi$ tail of the distribution. It is therefore desirable to  perform a
measurement at small $\xi$ to get a good sensitivity to $\dkap$ but also to 
be sensitive to events at high $\xi$ to observe the effect of the $\lam$ anomalous parameter.
Let us also note that it is useful to get an increased acceptance at high $\xi$ to benefit from
the large increase of the cross section when $\lam$ is different from 0 since this enhancement
appears only at high $\xi$.  In the following, we will distinguish two cases: either we perform a
``counting" experiment cutting only on the $\xi$ variable to enhance the effect of the anomalous
coupling, or we use angular variables to allow to distinguish with the SM predictions. 

\par The current best limits on anomalous couplings come from the combined fits of 
all LEP experiments~\cite{LEPlimits}.
\begin{equation}
   -0.098<\dkap<0.101 \quad -0.044<\lam<0.047.
\end{equation}
The CDF collaboration presented the most stringent constraints on $\wwgamma$
coupling measured at hadron colliders~\cite{TEVlimits}  
\begin{equation}
   -0.51<\dkap<0.51 \quad  -0.12<\lam<0.13
\end{equation}
analyzing the $W\gamma$ events in parton-parton interactions.
Even though the LEP results are more precise than the results from the hadron 
collider, there is always a mixture of $\gamma$ and $Z$ exchange present in the
process $e^+e^-\rightarrow WW$ from which the couplings are extracted. The 
two-photon $WW$ production have the advantage that pure $W-\gamma$ couplings are tested
and no $Z$ exchange is present.

\subsection{Sensitivity to anomalous parameters with the counting experiment}

The sensitivity to anomalous coupling can be derived by counting the number
of observed events and comparing it with the SM expectation. The statistical
significance is defined as $|\Delta N|/\sqrt{N_{SM}+N_{\pomeron}}$, where $\Delta N$
is the difference between the number
of events predicted by the SM Lagrangian $N_{SM}$ and the Lagrangian with
nonzero anomalous parameters, and $N_{\pomeron}$ denotes the number of events due to DPE
background. This assumes that the background ---
pure diphoton SM and DPE --- follows a Gaussian distribution. We consider two 
running scenarios with different acceptances of the forward detectors for
AFP-CMS (AFP stands for ATLAS Forward Physics) and CMS-TOTEM experiments to derive the sensitivity on anomalous parameters:
\begin{itemize}
   \item AFP-CMS - standard running condition of ATLAS or CMS/TOTEM forward detectors at
   220 and 420 m with 
an acceptance of $0.0015<\xi<0.15$, which we mentioned already 
   \item CMS-TOTEM - running with forward detectors around the CMS interaction point 
at 420\units{m} and in addition the detectors of TOTEM experiment at 147\units{m} and 220\units{m}
with an overall acceptance of $0.0015 < \xi <0.5$ (based on the Ref. \cite{TOTEMnote}) as a mean
to reach higher sensitivity~\footnote{Note that the study using the TOTEM acceptance
up to $\xi=0.5$ should be seen as an approximate investigation of the physics
potential since the precise determination of the TOTEM acceptance
is still ongoing.} to $\lam$.
\end{itemize}
The TOTEM forward detectors are placed closer to the interaction point
and hence have access to higher $\xi$. This is very desirable to enhance
the sensitivity to $\lam$ since the $\lam$ signal manifests itself in the 
region of high $\xi$ as we mentioned in the previous section (see Fig.~\ref{fig:xianom}). 

\par In order to obtain the best $S/\sqrt{B}$ ratio, the
$\xi_{min}<\xi_i<\xi_{max}$ acceptance was further optimized for the $\lam$ parameter.
The event is accepted if $\xi_i>0.05$ for the AFP-CMS scenario and if $\xi_i>0.1$
for the CMS-TOTEM scenario. These optimization cuts do not change significantly
between studies with and without coupling form factors. 
In case of $\dkap$, the full acceptance of the AFP detectors is used since 
the difference between the enhanced and SM cross section
is almost flat around relevant values of the coupling $|\dkap|\sim0.02$, see
Fig.~\ref{fig:xianom} (left). To summarize, the corresponding acceptance 
cuts used to derive limits on the coupling parameters are shown in Table \ref{tab:optim}.
\begin{table}[htb]
\begin{tabular*}{.3\textwidth}{@{\extracolsep{\fill}}l|cc}
         &  $\xi_{min}$   &   $\xi_{max}$ \\ \hline
\hline
$\dkap$ with AFP-CMS  & 0.0015        &   0.15        \\
$\lam$  with AFP-CMS &  0.05          &   0.15        \\
\hbox{$\lam$ with CMS-TOTEM}      &  0.1            & 0.5              
\end{tabular*}
\caption{Different acceptances of the forward detectors used to derive the limits on TGCs for two
running scenarios: AFP-CMS detectors and CMS-TOTEM forward detectors.}
\label{tab:optim}
\end{table} 

\par
\begin{figure}[hbt]
\includegraphics[width=1\picwidth]{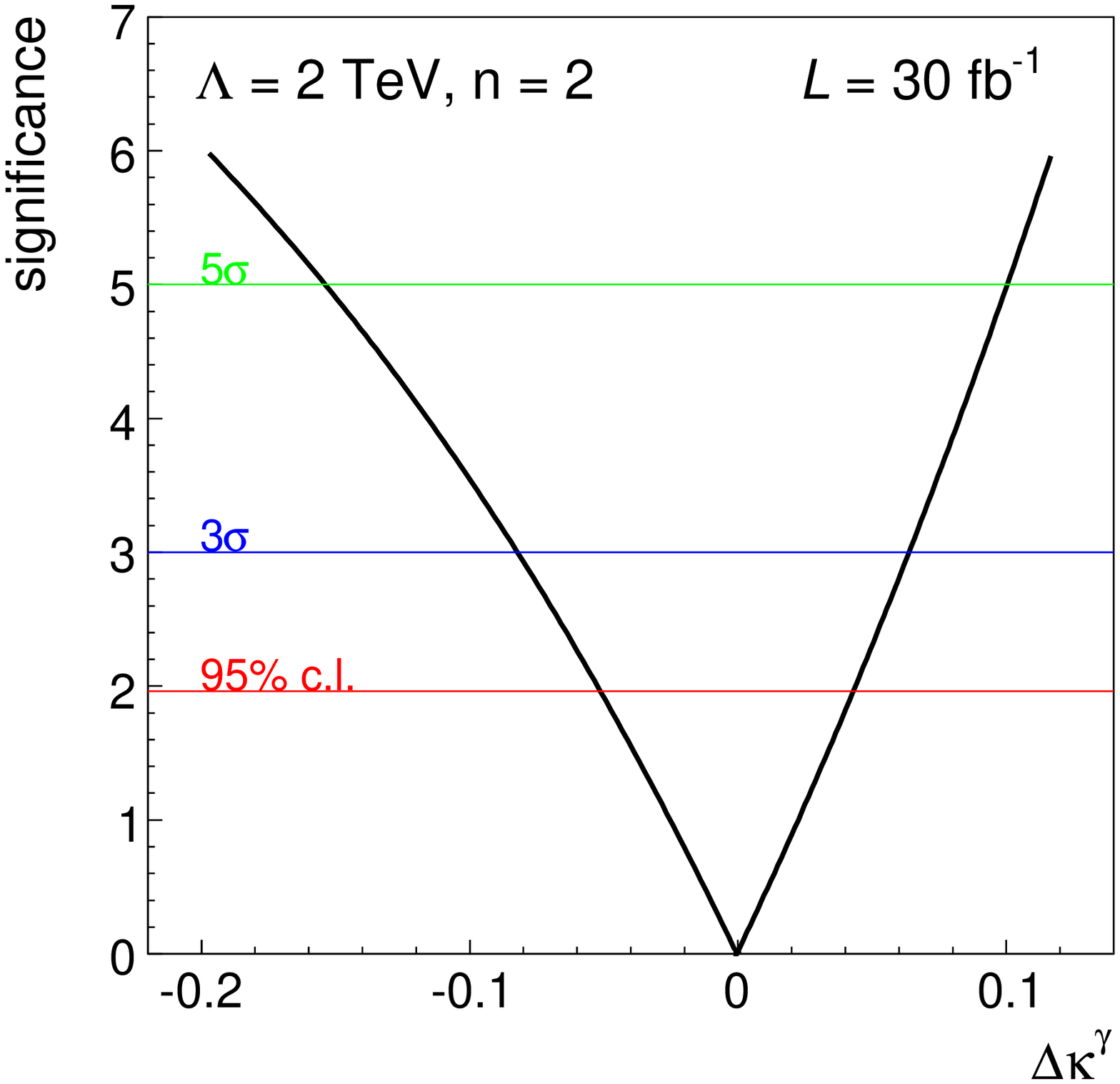}
\includegraphics[width=1\picwidth]{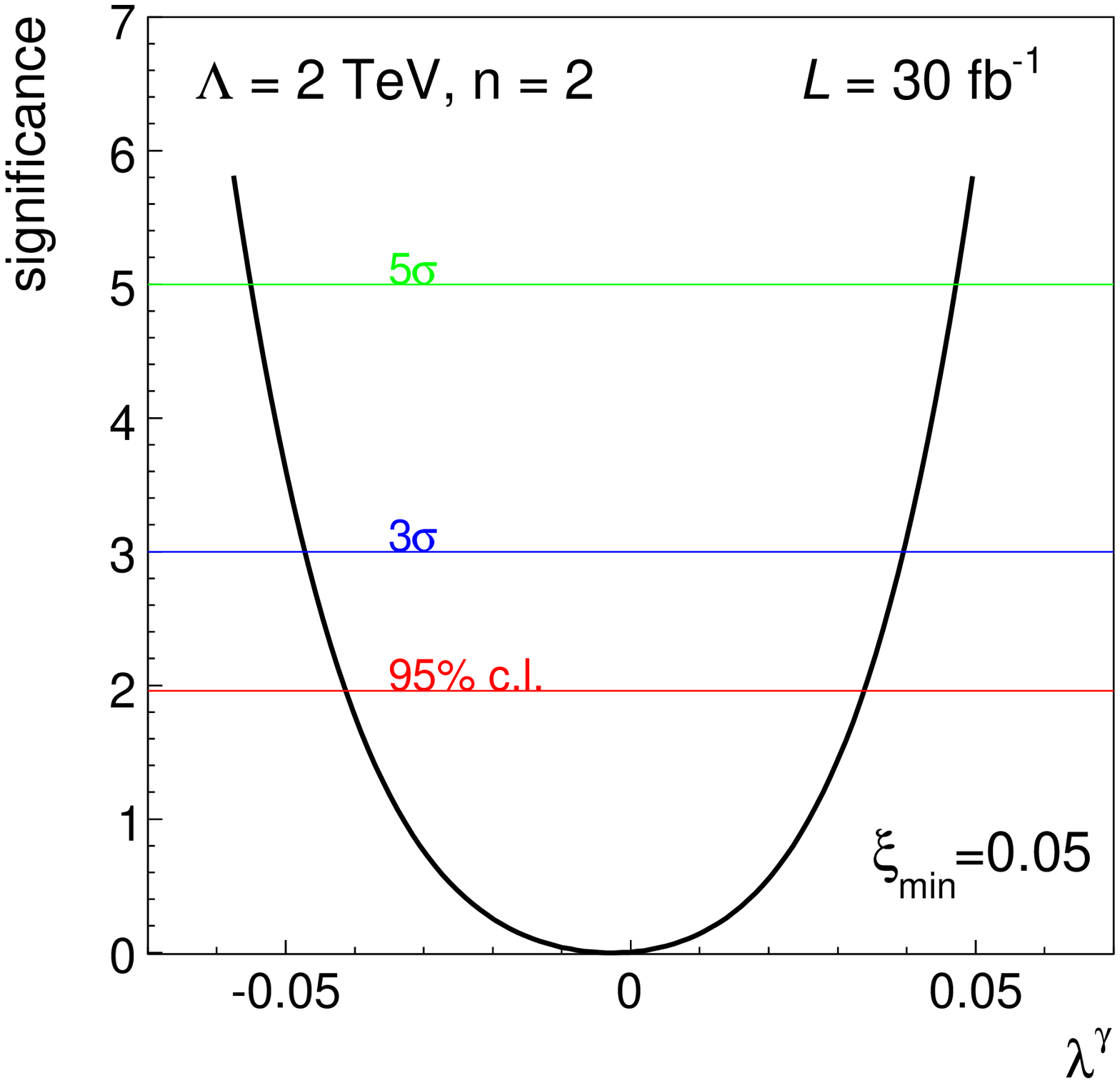}\\
\includegraphics[width=1\picwidth]{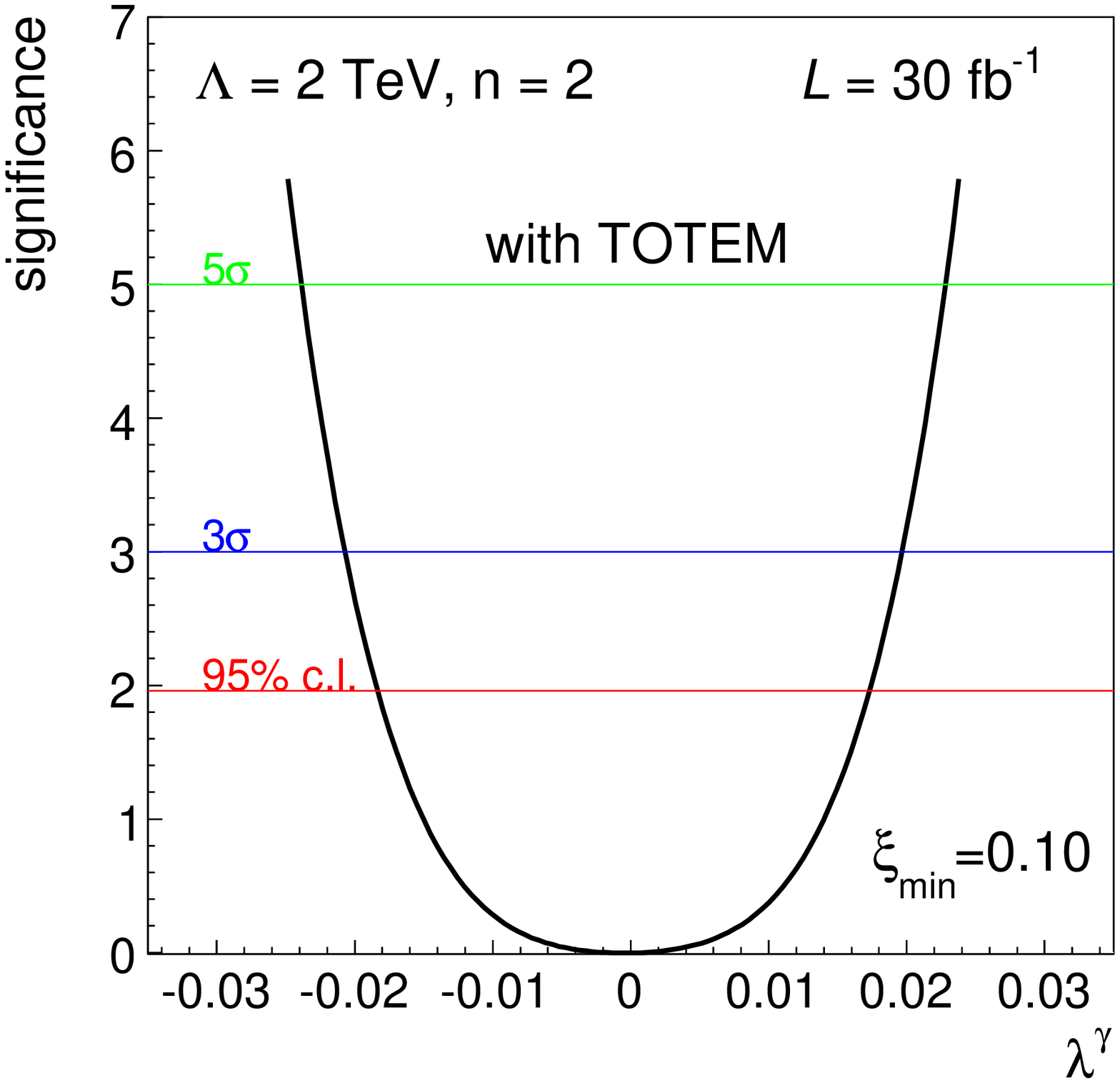}
\caption{Significances as a function of the anomalous parameters $\dkap$ (left) and $\lam$
(right) for \lumi=30$\invfb$ using the AFP-CMS forward detectors. The dependence on the 
anomalous parameter $\lam$ using the CMS-TOTEM acceptance is also shown (bottom plot). 
The anomalous parameters were multiplied by the form factors (see equation~(\ref{eq:formfactors})).
$\xi_{min}$ is the minimum momentum fraction loss of the proton used to enhance the S/$\sqrt{B}$ ratio.}
\label{fig:limits30}
\end{figure}

 The dependence of the significance on the anomalous parameters multiplied by the
form factors and calculated for an integrated luminosity of $30\invfb$ is
shown in Fig.~\ref{fig:limits30} for the AFP-CMS and CMS-TOTEM scenarios.
The 95\,\% c.l. limits,
$3\sigma$ evidence and $5\sigma$ discovery are indicated.  The corresponding
$3 \sigma$ evidence, $5\sigma$ discovery and 95\% confidence level limits derived for a 
luminosity of \lumi=$30\invfb$ 
are shown in Table \ref{tab:limits30} together with the limits
obtained without form factors taken into account. 
As was mentioned above, the sensitivity to $\lam$ is about two times better
using the TOTEM roman pots at 147 m. The number of signal and background events is also
given in  Table \ref{tab:eventsLimits30} for a 95\% c.l. limit.
\begin{table}[h]
\begin{tabular}{r}
 \\
  95\% c.l           $\Bigg\{$  \\
  $3\sigma$ evidence $\Bigg\{$\\
  $5\sigma$ discovery $\Bigg\{$\\
\end{tabular}
\begin{tabular}{l|ccc}
form factor             &  $\dkap$(AFP)   &    $\lam$(AFP)      &    $\lam$ (CMS+TOTEM)     \\
\hline
$ \Lambda=\infty$            &   [-0.034, 0.029]  &  $[-0.033,0.026]$  &    $[-0.015, 0.014]$ \\
$ \Lambda=2\units{TeV}$      &   [-0.051, 0.043]  &  $[-0.041,0.034]$  &    $[-0.018, 0.017]$ \\    
\hline
\hline
$ \Lambda=\infty$            &   [-0.053, 0.044]  &  $[-0.038, 0.031]$  &    $[-0.017, 0.016]$ \\
$ \Lambda=2\units{TeV}$      &   [-0.082, 0.064]  &  $[-0.047, 0.040]$  &    $[-0.021, 0.020]$ \\ 
\hline
$ \Lambda=\infty$            &   [-0.097, 0.069]  &  $[-0.047, 0.038]$  &    $[-0.019, 0.018]$ \\
$ \Lambda=2\units{TeV}$      &   [-0.154, 0.100]  &  $[-0.055, 0.047]$  &    $[-0.024, 0.023]$ \\ 
\end{tabular}
\caption{95\% c.l., 3$\sigma$ evidence, and $5\sigma$ discovery potential
on the $\wwgamma$ anomalous parameters using AFP-CMS or CMS-TOTEM forward detectors 
with and without form factors applied for a luminosity of
\lumi=$30\invfb$.}
\label{tab:limits30}
\end{table}

\begin{table}[h]
\begin{tabular}{c|ccc}
form factor             &  $\dkap$(AFP)   &    $\lam$(AFP)      &    $\lam$ (CMS+TOTEM)     \\
\hline
$\Delta N$            &   60       &  15      &  25    \\
$N_{SM}$              &   842      &  23      &  3      \\
$N_{\pomeron}$      &   65     &  27      &  152     \\
\hline
$95\%$ c.l.     &   [-0.051, 0.043]  &  $[-0.041,0.034]$  &    $[-0.018, 0.017]$ \\    
\end{tabular}
\caption{Number of signal events $\Delta N$, SM events $N_{SM}$, and
background $N_{\pomeron}$ to be observed for the values of anomalous couplings
corresponding to the 95\% c.l. limits for $\dkap$ and $\lam$ (bottom line) in AFP-CMS and
CMS-TOTEM running scenarios.}
\label{tab:eventsLimits30}
\end{table}

\par To study the best possible reach on measuring the anomalous parameters
using the AFP-CMS detectors we derive the limits also for a luminosity of \lumi=$200\invfb$. 
The significances were calculated in the same way as before and 
the results are summarized in Table~\ref{tab:limits200}. We note that the present sensitivities
coming from the Tevatron experiments
can be improved by about a factor 30, while the LEP sensitivity can be improved
by a factor 5. 
\begin{table}[h]
\begin{tabular}{r}
 \\
  95\% c.l           $\Bigg\{$  \\
  $3\sigma$ evidence $\Bigg\{$\\
  $5\sigma$ discovery $\Bigg\{$\\
\end{tabular}
\begin{tabular}{l|ccc}
form factors               &$\dkap$(AFP)  &   $\lam$(AFP)         &    $\lam$ (CMS+TOTEM)     \\
\hline
$ \Lambda=\infty$          & [-0.013, 0.012]     &  [-0.024, 0.017] &    $[-0.011,0.010]$       \\
$ \Lambda=2\units{TeV}$    & [-0.019, 0.017]     &  [-0.030, 0.023] &    $[-0.014,0.013]$      \\
\hline
\hline
$ \Lambda=\infty$            &   [-0.019, 0.018]  &  $[-0.028, 0.021]$  &    $[-0.013, 0.012]$ \\
$ \Lambda=2\units{TeV}$      &   [-0.029, 0.026]  &  $[-0.035, 0.028]$  &    $[-0.016, 0.015]$ \\ 
\hline
$ \Lambda=\infty$            &   [-0.033, 0.029]  &  $[-0.033, 0.026]$  &    $[-0.015, 0.014]$ \\
$ \Lambda=2\units{TeV}$      &   [-0.051, 0.042]  &  $[-0.041, 0.034]$  &    $[-0.018, 0.017]$ \\ 
\end{tabular}                                            
\caption{95\% c.l., 3$\sigma$ evidence, and $5\sigma$ discovery potential on the $\wwgamma$ 
anomalous parameters for a luminosity of \lumi=200$\invfb$
using the AFP-CMS or TOTEM-CMS detectors.}
\label{tab:limits200}
\end{table}

\par
\begin{figure}
   \includegraphics[width=\picwidth]{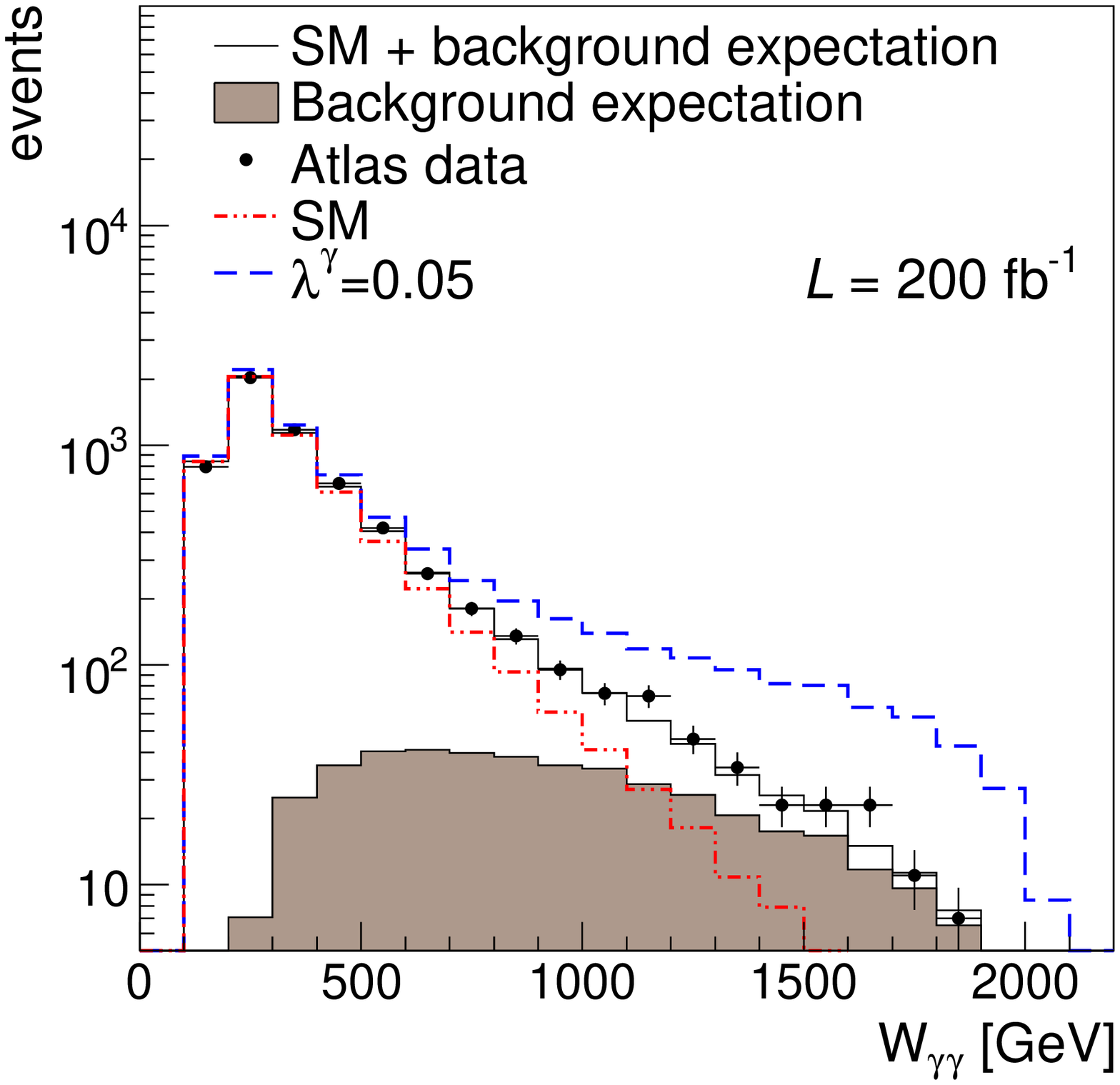}
   \includegraphics[width=\picwidth]{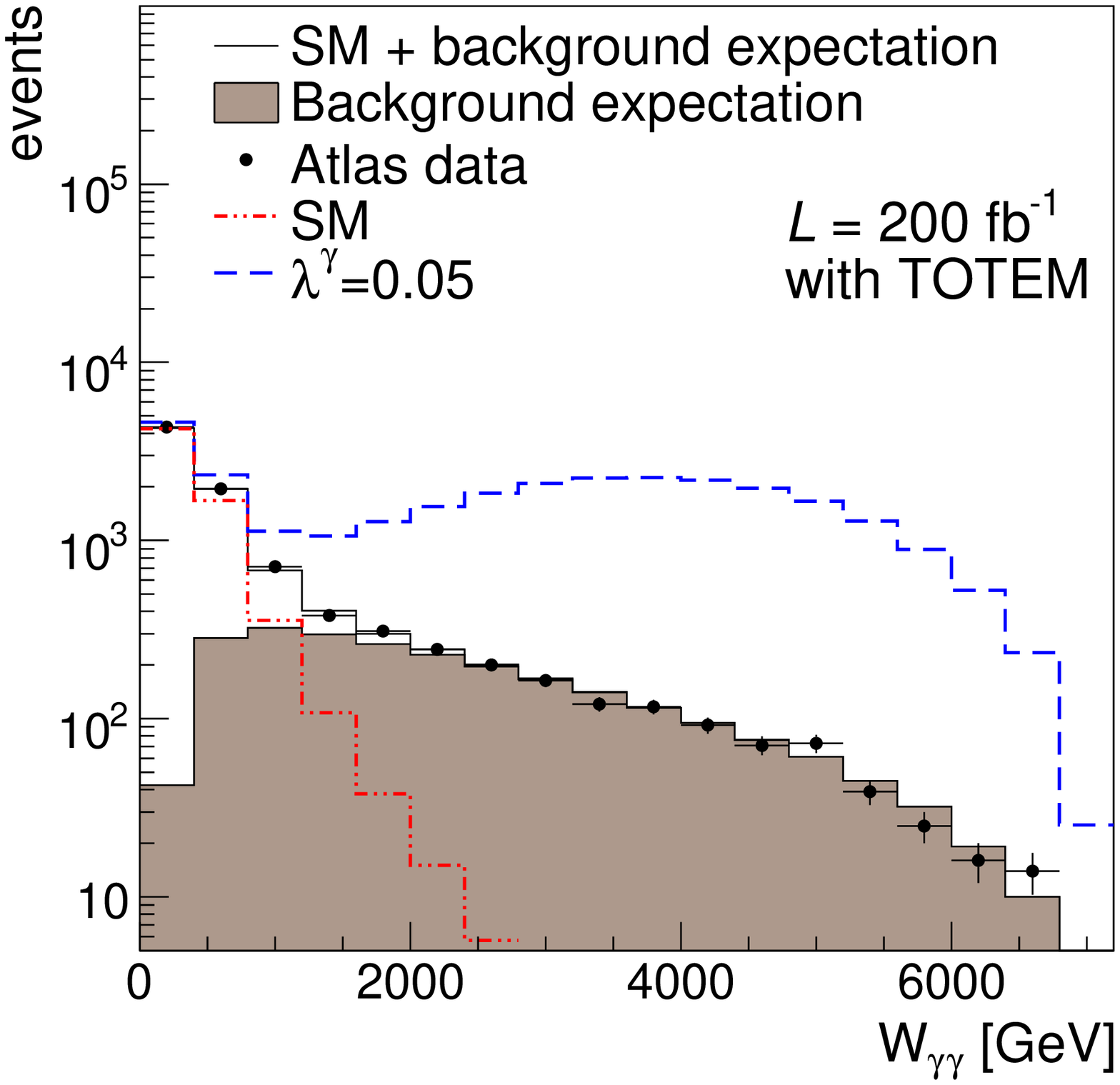}
\caption{Distributions of the $\gamma\gamma$ photon invariant mass $W_{\gamma\gamma}$ 
measured with the forward detectors using $W_{\gamma\gamma}=\sqrt{s\xi_1\xi_2}$
for AFP-CMS (left)
and CMS-TOTEM (right). The 
effect of the $\lam$ anomalous parameter appears at high $\gamma\gamma$ 
invariant mass (dashed line).  The SM background is indicated in dot-dashed line, 
the DPE background as a shaded area and their combination
in full line. The black points show the ATLAS data smeared according to a Poisson distribution. The signal due to the anomalous coupling $\lam$ is better
visible when the full acceptance of TOTEM up to $\xi=0.5$ is used (right figure).}
\label{fig:Wspectrum200}
\end{figure}

 A similar observation can be made using the distribution of the $\gamma\gamma$ invariant mass.
After tagging the protons with forward detectors, the invariant mass can be calculated as $W_{\gamma\gamma}=\sqrt{s\xi_1\xi_2}$
with a resolution of 2-3 GeV.
The signal due to the anomalous parameter $\lam$ appears at high 
$W_{\gamma\gamma}$ and induces a 
change in shape of the spectrum, see Fig.~\ref{fig:Wspectrum200}.
At low $W_{\gamma\gamma}$, 
the dominant background to the anomalous signal is the SM two-photon production,
whereas
at high masses, the SM signal is very small and the background is mainly due to the DPE events. 
The situation is more favorable in case of CMS-TOTEM setup which has an acceptance to higher 
fractional momentum loss ($\xi_{max}=0.5$)
and it is possible to tag higher $\gamma\gamma$ mass, see Fig.~\ref{fig:Wspectrum200} (right).
The sensitivity on anomalous coupling is similar to the one we obtained using
the counting experiment but cutting on $W$. The effect of the $\dkap$ anomalous coupling is to
change the normalisation of the $W$ distribution without modifying the shape.
It means that the measurement of the $W$ differential cross section is a way
to distinguish between $\lam$ and $\dkap$ couplings as well. It is also worth
noticing that a binned likelihood fit of the $d \sigma/dW$ shape of the
distribution will allow to improve the sensitivity of $\lam$ described in this
paper once the forward detectors and their acceptance parameters will be
well understood to distinguish between detector and physics effects.  

\subsection{$p_T^l$ and angular distributions using an integrated luminosity of
\lumi=200$\invfb$}

\par In this section we study some other observables related to angular distributions
sensitive to anomalous parameters. However, it is needed to collect enough luminosity
to be sensitive to a particular shape of the distributions and this is why we will
study these distributions only for the highest luminosity, \lumi=200$\invfb$.

\par
\begin{figure}
      \includegraphics[width=\picwidth]{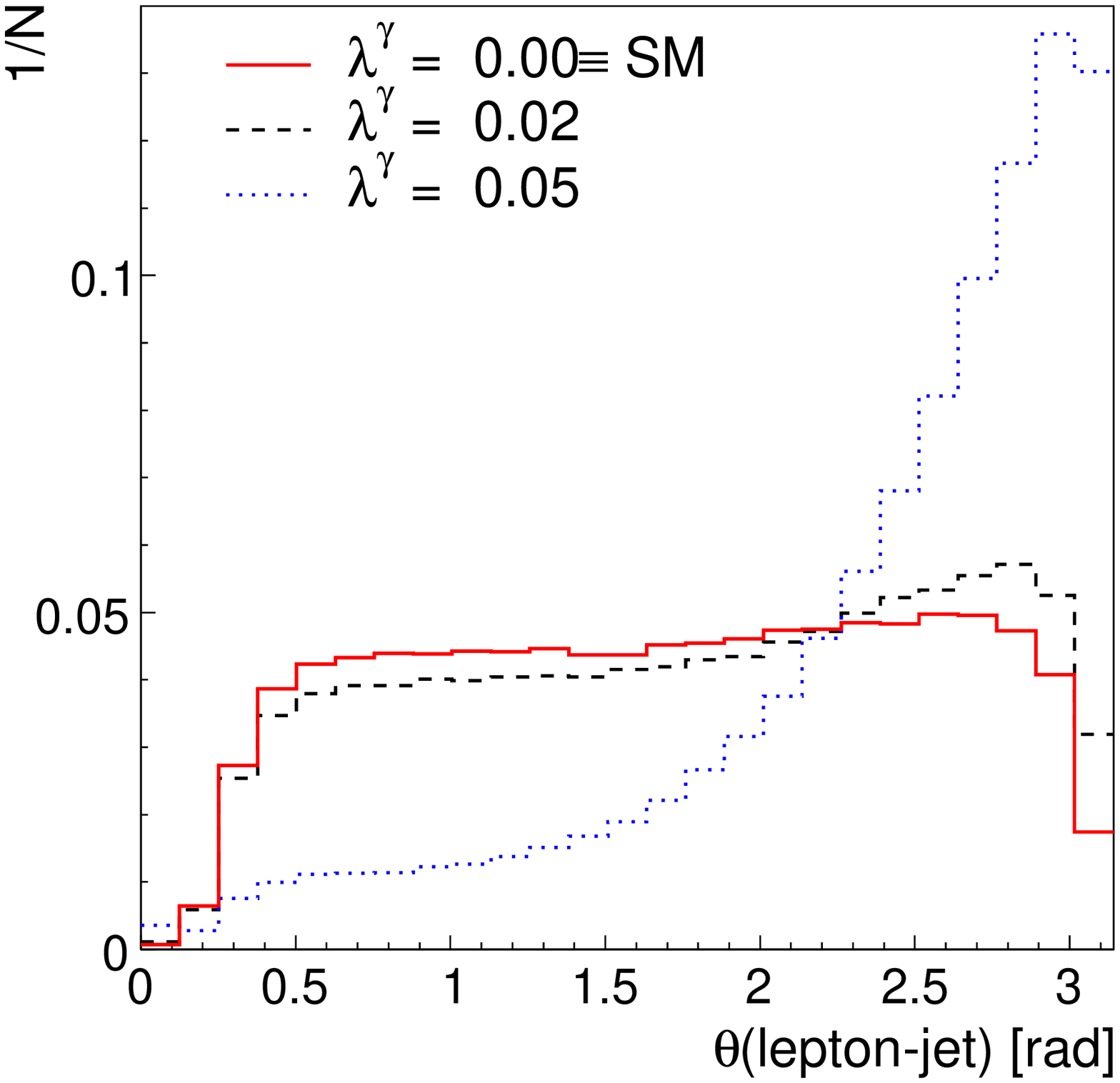}
      \includegraphics[width=\picwidth]{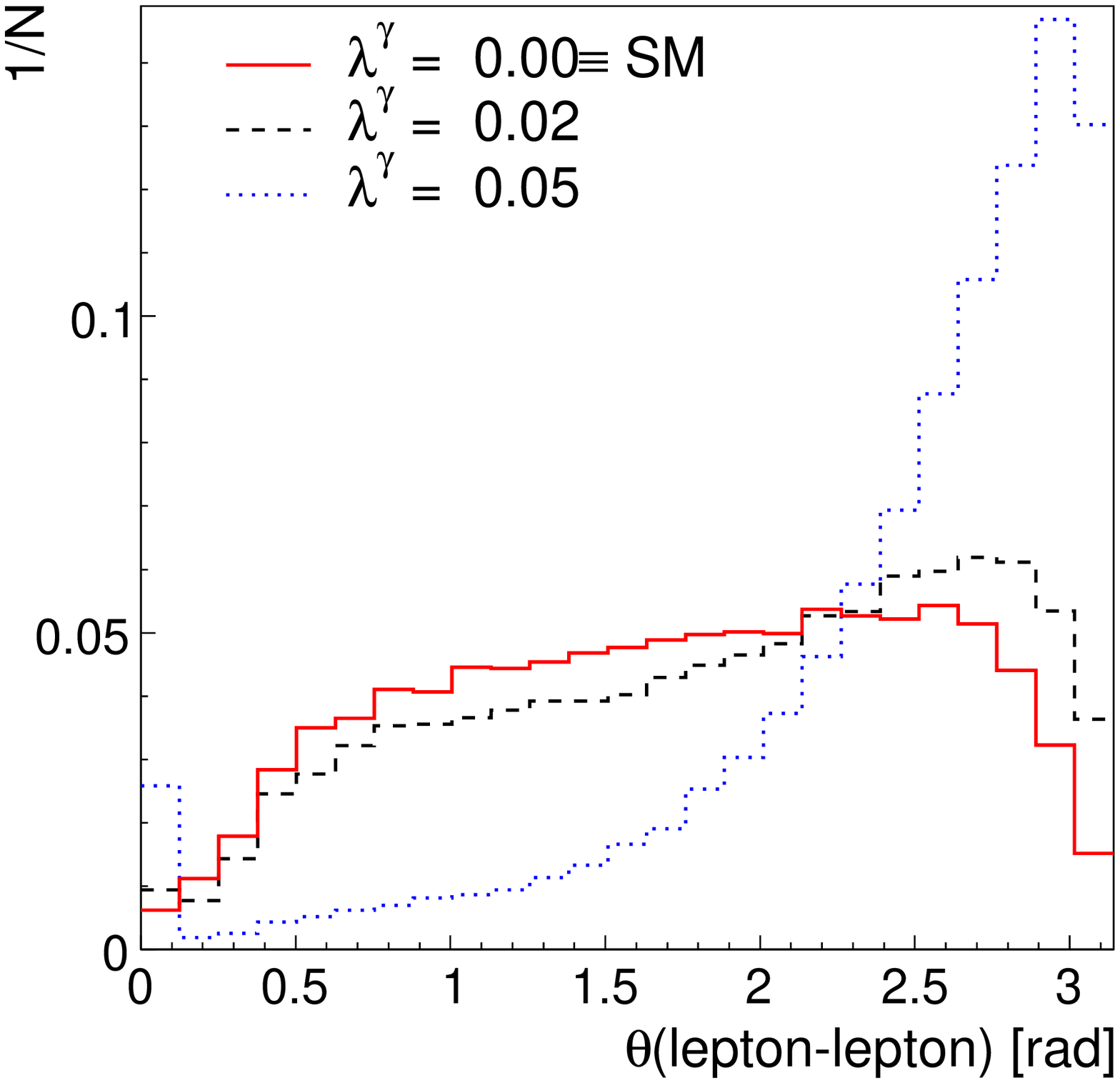}\\
      \includegraphics[width=\picwidth]{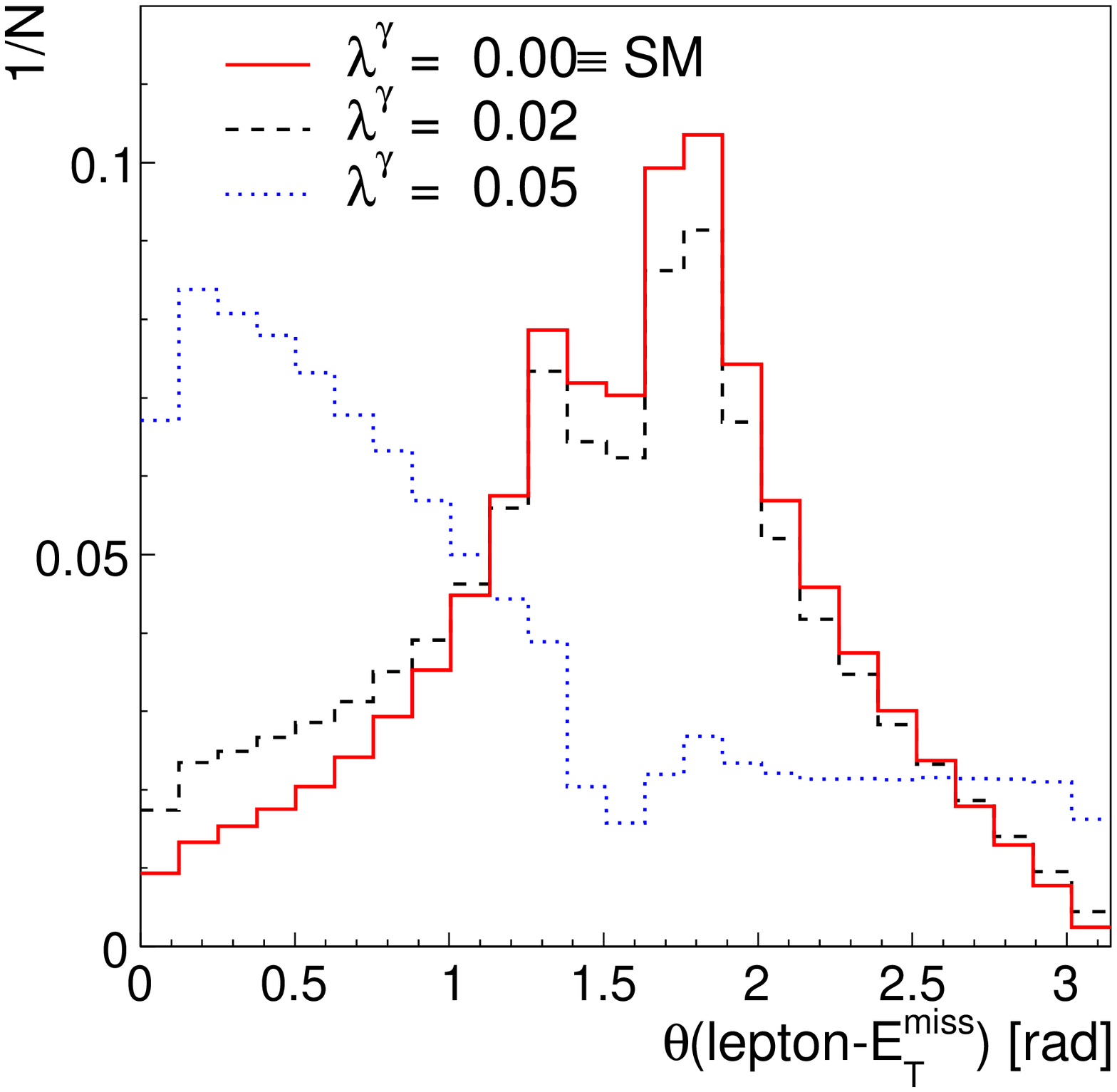}
\caption{Various angular distributions for $\lam=0, 0.02, 0.05$:
angle between the leading lepton and the leading jet in the event (top left), angle between the 
two leading leptons in the event (top right), and angle between the leading lepton and the missing transverse
energy at generator level. 
The shape of the angular distribution changes as a function of $\lam$  (curves 
for negative values of $\lam$ are not shown because they are similar to the ones with positive $\lam$). 
Note that the shape of the angular distributions does not change with $\dkap$ and is therefore not shown here.
}
\label{fig:angularGen}
\end{figure}

 Angular distributions computed at generator level are displayed 
in Fig.~\ref{fig:angularGen}. The
distribution of the angle between the leading lepton and the leading jet, the
angle between the leading and the sub-leading lepton
(ordered in $p_T$), and the angle between the leading lepton and the vector of the 
missing energy in the event are shown.
All shapes are significantly different when $\lam$ is shifted from the SM value. 
We note that concerning the $\dkap$ parameter, the shape of the angular distributions is modified 
negligibly  and cannot be used to derive better sensitivities by fitting the shape of the distributions.

\par
\begin{figure}
   \includegraphics[width=\picwidth]{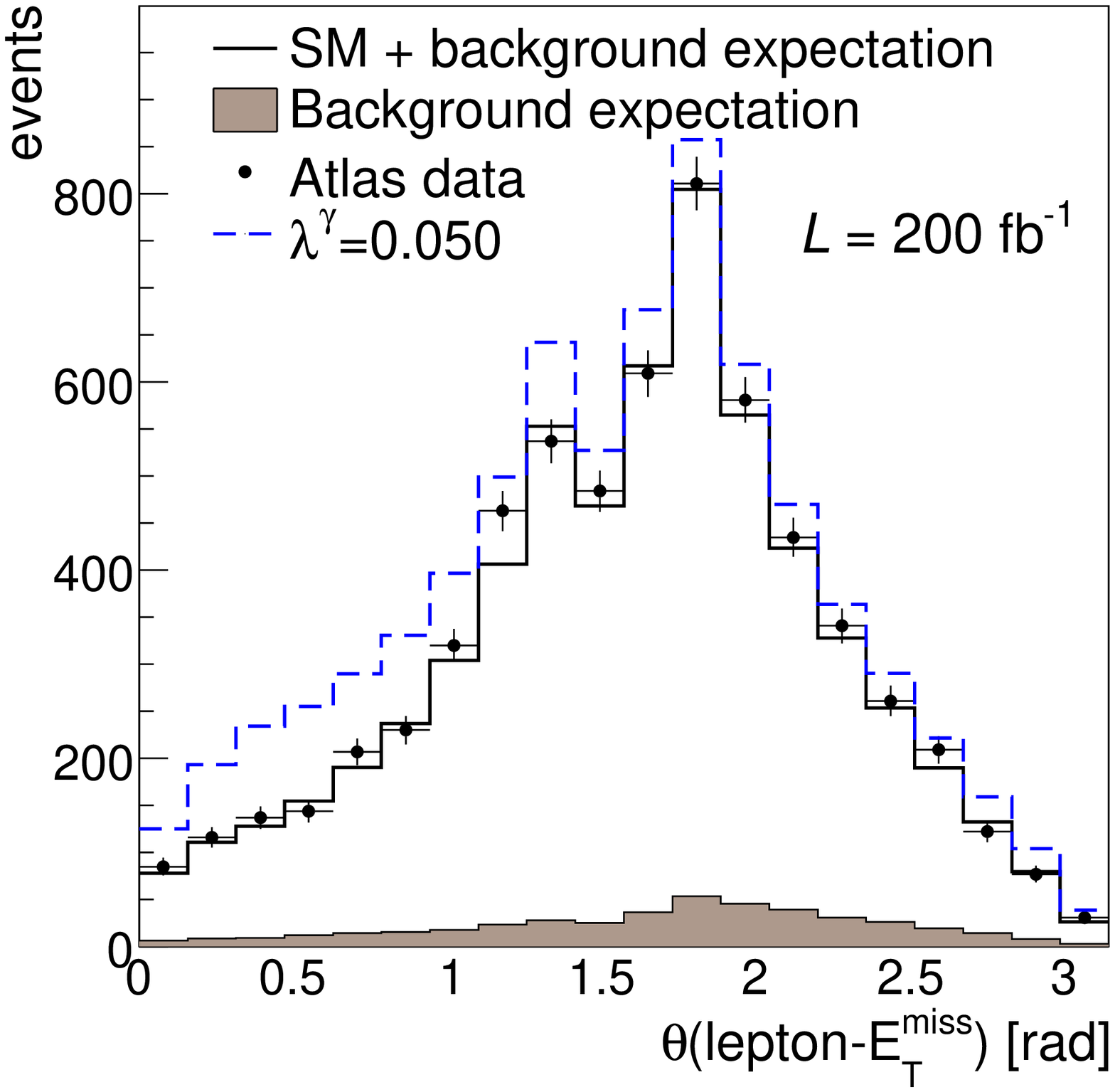}
   \includegraphics[width=\picwidth]{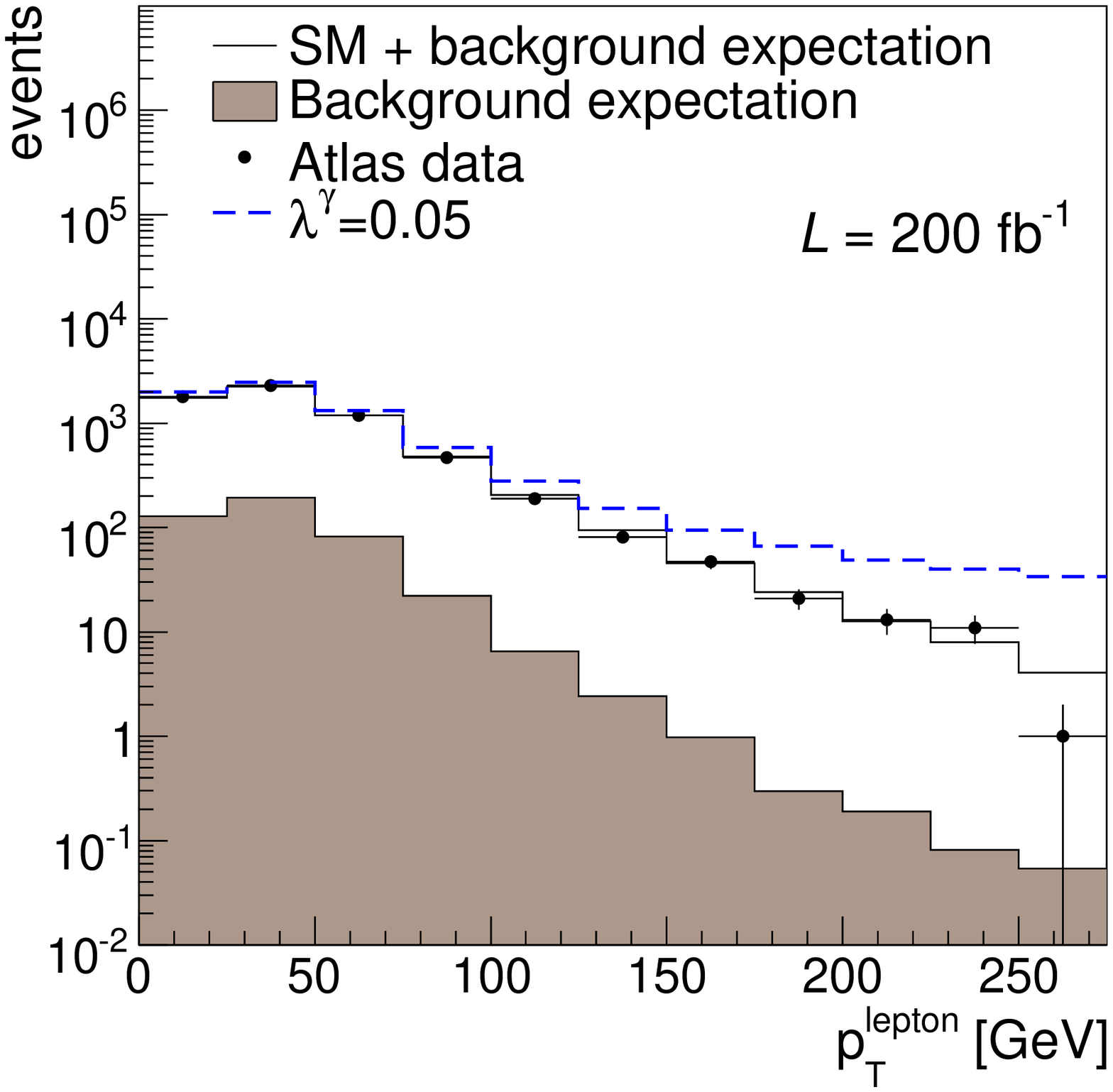}
\caption{\textit{Left}: distribution of the angle between the leading lepton and $E_T^{miss}$. \textit{Right}:  
distribution of the leading lepton transverse momentum for \lumi=$200\invfb$.
The DPE background is given in the shaded area ( it is one order of magnitude smaller
than the SM expectation), the SM and DPE background in full line and the 
effect of an anomalous coupling $\lam=0.05$ in dashed line. The 
black points show the ATLAS data smeared according to a Poisson distribution.}
\label{fig:distributions200}
\end{figure}

 The effect of the forward detector acceptance is such that the enhancement due to anomalous coupling
is suppressed in the region where $\lam$ enhances the cross section. 
For example, in Fig.~\ref{fig:distributions200} the angular distribution of the leading 
lepton-$E_T^{miss}$ angle is shown for the DPE background, the SM and DPE background 
and the effect of anomalous coupling. Even though some enhancement is seen for a value of $\lam=0.05$, 
it is much smaller than at generator level (Fig.~\ref{fig:angularGen}).  In addition, the distribution of
the transverse momentum of the leading lepton is presented in Fig.~\ref{fig:distributions200}. The signal
due to the anomalous parameters appears in the region of high lepton transverse momentum. 

\par Unfortunately, the forward detector acceptance does not allow to benefit fully from the shape difference between
the SM model background and the signal, and this is why it is not easy to improve the sensitivities
reached by the counting experiments described in the previous section.
However, the DPE background is by one order of magnitude smaller
than the two-photon $W$ pair production and testing the shape of the observed distribution 
against the Monte Carlo prediction on the basis of the binned log-likelihood method could lead to some improvement on the $\lam$ sensitivity.

\subsection{Comparison with other LHC measurements}

In this section, we compare our results of the 95\% c.l. limits with the
standard methods used for instance in ATLAS~\cite{SMnoteMaarten} and
\cite{DobbsNote}, to determine if our new method to assess anomalous coupling
is competitive at the LHC. Expectations for the CMS collaborations are found to
be similar.

\par In the inelastic channel, the $\wwgamma$ anomalous coupling is probed by fitting the 
$p_T^{\gamma}$ of the photon distribution to the NLO expectations using the combined sample
of $W(e\nu)\gamma$ and $W(\mu\nu)\gamma$ events or by fitting the transverse mass
distribution $M_T(WW)$ of the boson pair in events with two $W$s in the final state.
The corresponding 95\% c.l. limits obtained for \lumi=30$\invfb$ assuming $\Lambda$=2\TeV 
and $n=2$ for the coupling form factors are shown in Table \ref{tab:LHCcl}. The current
analysis using the forward detectors in ATLAS has about the same precision as the analysis
in the inelastic channel in ATLAS and is therefore complementary 
to that performed without tagging the forward protons (see Table
\ref{tab:limits30}). In addition, the anomalous parameter $\lam$ can be
event better constrained if the $\xi$ acceptance is larger as is in the CMS-TOTEM running scenarios
mentioned above. In this case one gains about a factor 2-5 higher precision than in the conventional ATLAS
analysis. Let us note also that the standard ATLAS analysis suffers from the difficult $\gamma$ selection and
energy scale whereas the presence of the
forward detectors allows to perform an easier counting experiment.

\begin{table}[htb]
\begin{tabular}{ccc}
                         &     $\dkap$        &  $\lam$         \\
$W\gamma, (p_T^{\gamma})$&     [-0.11, 0.05]  &  [-0.02, 0.01]  \\
$WW, (M_T)$              &     [-0.056, 0.054]&  [-0.052. 0.100]  \\
\end{tabular}
\caption{95\% c.l. limits on the $\wwgamma$\ coupling parameters obtained from fitting the
$p_T^{\gamma}$ and $M_T(WW)$ distributions in $W\gamma$ and  $WW$ final states in inelastic production
in ATLAS, and calculated for \lumi=$30\invfb$ and for the form factors $\Lambda=2$\TeV, $n=2$.
\cite{SMnoteMaarten}. }
\label{tab:LHCcl}
\end{table}

\section{Conclusion}
In this paper, we first discussed a new possible test of the SM by measuring
the $WW$ production via photon induced processes which has a cross section 95.6 fb. This measurement assumes the
detection of the decay product of the $W$ in the main central detectors of ATLAS
or CMS and the presence of forward detectors allowing to measure the intact
scattered proton after the interaction. To remove most of the QCD background,
only the cases when at least one of the $W$ decays leptonically (electron or 
muon) is considered. After these assumptions, the cross section is of the order of 28.3 fb, and most of the
dominant background due to DPE can be removed by a cut on $\xi<0.15$.
With a low LHC luminosity of 200 pb$^{-1}$, it is possible to observe a signal
of 5.6 $WW$ events with a low background less than 0.4 events, leading to a
signal above 8.6 $\sigma$. The measurement of the cross section can be compared
to the precise QED expectation from the SM.

In a second part of the paper, we described how this measurement is sensitive
to anomalous $WW\gamma$ coupling. We considered the modification of the
$WW\gamma$ vertex with additional terms conserving $C$ and $P$ parity only, that
are parametrized with two anomalous parameters $\dkap$ and $\lam$. 
The advantage with respect to LEP is that we are sensitive only to the
$WW\gamma$ coupling and not to the $WWZ$ one. A simple
counting experiment measuring the number of $WW$ photon induced events and
cutting on the $\xi$ of the protons in the final state allows to gain about a
factor 30 on anomalous couplings with respect to present sensitivity
from the Tevatron, and a factor 5 compared to LEP results
after accumulating 200 fb$^{-1}$. Typically, the measurement
extends to low $\xi$ (down to 0.0015) and to high $\xi$ (up to 0.15-0.5) to
obtain the best sensitivity to  $\dkap$ and $\lam$, respectively. 
The best sensitivity on $\dkap$ and $\lam$  is respectively 
$[-0.013,0.012]$ and $[-0.011,0.010]$. Analyzing angular distributions at the
LHC for instance between the leading lepton and the leading jet or the leading
lepton and the sub-leading lepton only improves marginally the results since the
region where most of the difference in shape appears is cut off by the forward
detector acceptance.

Last but not least, it is worth mentioning that studying the differential $WW$
production cross section via photon induced processes as a function of $W$ is
sensitive to beyond standard model effects (SUSY, new strong dynamics at the TeV
scale, anomalous coupling, etc.) for $W \sim$ 1 TeV. It is expected that the LHC
experiments will collect 400 such events predicted by QED with $W>$1 TeV for a luminosity of 200
fb$^{-1}$ which will allow to probe further the SM expectations. In the same way
that we studied the $WW\gamma$ coupling, it is also possible to study the
$ZZ\gamma$ one. The SM prediction for the $ZZ\gamma$ coupling is 0, and any
observation of this process is directly sensitive to anomalous coupling (the main
SM production of exclusive $ZZ$ event will be due to exclusive Higgs boson
production decaying into two $Z$ bosons).
 
\section*{Acknowledgments} 
The authors thank K.~Piotrzkowski and M.~Boonekamp for motivating this study
and many discussions about it and in addition
A.~Kup\v{c}o,
F.~Chevallier, C.~Marquet and R.~Peschanski for reading the manuscript.

\section*{Appendix}
\appendix
In this section we provide a list of formulae that were used for the
calculation  of the $W$ pair production and are referenced in the text.
\begin{itemize}

\item 
The photon radiation by the proton is described by the photon spectrum. The 
photon spectrum in the Equivalent Photon Approximation is integrated from 
a kinematic minimum $Q^2_{min}$ up to $Q^2_{max}$ as a function of the photon 
energy $E_{\gamma}$ and reads \cite{Budnev}
\begin{equation}
\d N(\egamma)=\frac{\alpha}{\pi}\frac{\d\egamma}{\egamma}
\lr{1-\frac{\egamma}{E}}
\left[\varphi\lr{\frac{Q^2_{max}}{Q^2_0} }-\varphi\lr{\frac{Q^2_{min}}{Q^2_0} }\right],
\label{app:eq:budnev}
\end{equation}

\begin{equation}
\varphi(x)=(1+ay)
\left[%
-\ln(1+x^{-1})+\sum_{k=1}^{3}\frac{1}{k(1+x)^{k}}
\right]
\oplus\frac{(1-b)y}{4x(1+x)^3}
+c(1+\frac{y}{4})
\left[%
\ln\frac{1+x-b}{1+x}+\sum_{k=1}^{3}\frac{b^k}{k(1+x)^{k}}
\right],
\nonumber
\end{equation}

\begin{equation}
y=\frac{\egamma^2}{E(E-\egamma)}, 
\quad a=\frac{1}{4}(1+\mu_p^2)+\frac{4m_p^2}{Q^2_0}\approx 7.16,
\quad b=1-\frac{4m_p^2}{Q_0^2}\approx -3.96, 
\quad c=\frac{\mu_p^2-1}{b^4}\approx 0.028,
\nonumber
\end{equation}
where $\alpha$ denotes the Sommerfeld fine-structure constant, $E$ the energy 
of the incoming proton, and $Q^2_0=0.71$. $m_p$ is the proton mass and $\mu_p\approx 7.78$
represents the magnetic moment of the proton. The circled plus sign in front of the second
term corresponds to the fixed sign error of formula (D.7) in~\cite{Budnev}.

\item The leading order differential formula for the $\gamma\gamma\rightarrow WW$ process
is a function of the Mandelstam variables $s,t,u$ and the mass of the vector boson $W$
\cite{electroweakCorrections}

\begin{equation}
\frac{\d\sigma}{\d\Omega}=\frac{3\alpha^2\beta}{2s}\left\{1
-\frac{2s(2s+3M_W^2)}{3(M_W^2-t)(M^2_W-u)}
+\frac{2s^2(s^2+3M_W^4)}{3(M_W^2-t)^2(M_W^2-u)^2}
\right\},
\label{app:eq:wwprod}
\end{equation}
where $\beta=\sqrt{1-4M_W^2/s}$
is the velocity of the $W$ bosons.
\end{itemize}

\end{document}